\documentclass[fleqn,usenatbib]{mnras}

\usepackage{newtxtext,newtxmath}

\usepackage[T1]{fontenc}

\DeclareRobustCommand{\VAN}[3]{#2}
\let\VANthebibliography\thebibliography
\def\thebibliography{\DeclareRobustCommand{\VAN}[3]{##3}\VANthebibliography}

\usepackage{graphicx}	
\usepackage{amsmath}	
\usepackage{multirow}
\usepackage{capt-of}
\usepackage{amsmath}
\let\oldAA\AA
\renewcommand{\AA}{\text{\normalfont\oldAA}}

\title[Stellar activity forecast]{The need for a public forecast of stellar activity to optimise exoplanet radial velocity detections \& transmission spectroscopy}

\author[Lalitha Sairam]{
Lalitha Sairam\thanks{E-mail: l.sairam@bham.ac.uk} and 
Amaury~H. M. J.~Triaud
\\
School of Physics \& Astronomy, University of Birmingham, Edgbaston, Birmingham B15 2TT, UK\\
}

\date{Accepted 20 May 2022. Received YYY; in original form ZZZ}

\pubyear{2022}

\begin{document}
\label{firstpage}
\pagerange{\pageref{firstpage}--\pageref{lastpage}}
\maketitle

\begin{abstract}
Advances in high-precision spectrographs have paved the way for the search for an Earth analogue orbiting a Sun-like star within its habitable zone. 
However, the research community remains limited by the presence of stellar noise produced by stellar magnetic activity. These activity phenomena can obscure the detection of Earth-mass exoplanets and can create parasitic signals in transmission spectra.
In this paper, we outline the need for a public forecast of stellar activity, and produce a proof of principle. Using publicly available spectra we are able to forecast stellar minima several years ahead and reach a typical uncertainty on the timing of these minima of $\pm 0.5~\rm year$, similar to the precision reached on our own Sun's magnetic cycle. Furthermore, we use our toy model to show that knowing when to observe can improve the sensitivity of HARPS-North's Solar telescope to low mass planets by up to an order of magnitude, and we show that the majority of exoplanets selected for Early Release Science and Guaranteed Time Observations on the {\it James Webb} will be observed close or during stellar maxima, incurring a higher risk of stellar contamination. 
We finish our paper by outlining a number of next steps to create a public forecast usable by teams around the globe, by telescope time allocation committees, and in preparation for spacecrafts such as {\it Ariel}.
\end{abstract}




\section{Introduction}

Debates around detected exoplanets continue to flourish due to possible confusion between the radial velocity signal produced by exoplanets and the induced shift in radial velocity created by variations in stellar activity. 
Signals intrinsic to the star give rise to radial velocity variability occurring at a wide range of time scales from minutes \citep[e.g. asteroseismic p-modes;][]{Chaplin_2019} to hours \citep[e.g. stellar granulation][]{Delmoro_2004, Bastein_2013}. While signals created by stellar p-modes are efficiently averaged out by using longer exposure times \citep{Chaplin_2019, Dumusque_2011, VedadPiMen}, stellar granulation can add substantial noise to radial velocity time series because of a larger amplitude and longer timescales than p-modes (\citealt{schrijver_zwaan_2000, Kjeldsen_2005,VedadPiMen}). Granulation noise is often reduced by observing the target several times per night separated by several hours \citep[e.g.][]{Dumusque_2011}.
The more challenging stellar activity signals for the detection and characterisation of exoplanets are produced by active regions on stellar surfaces producing variability on time scales related to stellar rotation  \citep[typically 10s of days, e.g.][]{Dumusque_2011}.

Signals in radial velocity can be caused both by dark stellar spots and by bright plages/faculae (\citealt{meunier_2017}, and references therein). 
Additionally, most stars also experience variability on timescales of several months to years called magnetic activity cycles \citep{Lindegren_2003}. Over a full cycle, the number of active regions increases and decreases, leading to a periodic increase and decrease in stellar variability.  

Signals produced by stellar activity hinder exoplanet detection in two ways: by completely drowning out or by mimicking the radial velocity signals of genuine exoplanets (\citealt{Saar_1997}, \citealt{Queloz_2001}, \citealt{desort_2007}). 
While there is abundant literature describing how to distinguish a change in radial velocity produced by a planet from that produced by stellar activity, regularly stellar activity trumps observers.  
For example, the announcement of  planets orbiting $\alpha$ Centauri (\citealt{Dumusque_2012,hatzes_2013}), Kapteyn{\textquoteright}s star (\citealt{anglada_2014, Robertson_2015, anglada_2016}), and Barnard{\textquoteright}s star (\citealt{ribas_2018, lubin_2021}) were artefact of stellar activity.
More recently work has had successes modelling stellar activity using photometry to predict the effect of stellar activity on radial velocities \citep[e.g.][]{aigrain_2012}, or by using non-parametric models such as Gaussian Process \citep[e.g.][]{rajpaul_2015, lalitha_2019, mascereno_2020}. However, these methods tend to require sampling radial velocities at a very frequent cadence, shorter than the frequency of most activity effects. 
For example, new observing campaign of Proxima Centuari as part of guaranteed time on ESPRESSO by \citet{mascereno_2020} was densely sampled. Similarly the future observations with HARPS-3 will be sampled at frequencies shorter than the associated activity periods \citep{hall_2018}.

The Sun, has an activity cycle of 11 years, the solar chromospheric activity indicator log(R'$_{\mathrm{HK}}$) varies from -4.75 during maxima to -5.0 during minima \citep{Dumusque_2011}. The amplitude of radial velocity perturbation can vary between 40 cm s$^{-1}$  and 140 cm s$^{-1}$ during minimum and maximum solar activity phases \citep{meunier_2010}. Hence, having a forecast would offer the opportunity to mitigate activity signals by focusing data intakes {\it only} when the activity signal is reduced, and where inaccuracies in modelling are less noticeable.

Another aspect of exoplanetary research affected by stellar activity is the atmospheric characterisation of exoplanets, particularly transmission spectroscopy, which is affected by changes in the host star spectrum on timescales of a planetary orbit (\citealt{agol_2010}, \citealt{sing_2011}). The varying presence of flares, plages and starspots, etc, and their distribution on the surface  can modify a planet's transit depth. The planet-to-star radius can be underestimated by 10\% in visible wavelengths due to an unocculted starspot \citep{oshagh_2014, McCullough_2014}.

Furthermore, if unocculted stellar surface features (such as spot) are sufficiently cool they can introduce false molecular features (e.g. $\rm H_2O$) into an exoplanet's spectrum \citep{barstow_2015,rackham_2018}.  Contamination has been suspected for several planets such as TRAPPIST-1 \citep{zhang_2018}, WASP-19 b \citep{espinoza_2018} and K2-18 b \citep{barclay_2021} but with increasingly precise spectra from James Webb Space Telescope ({\it JWST}) and Atmospheric Remote-Sensing Infrared Exoplanet Large-survey ({\it Ariel}, \citealt{tinetti_2016}), this list is bound to increase and contamination set to become a major hindrance to interpret these spectra and extract robust molecular and elemental abundances. Here too they are notable efforts to mitigate these effects \citep[e.g.][]{wakeford_2019}, but a forecast can also inform when the most indicated time to collect data is. 

\begin{figure}
\includegraphics[width=0.5\textwidth]{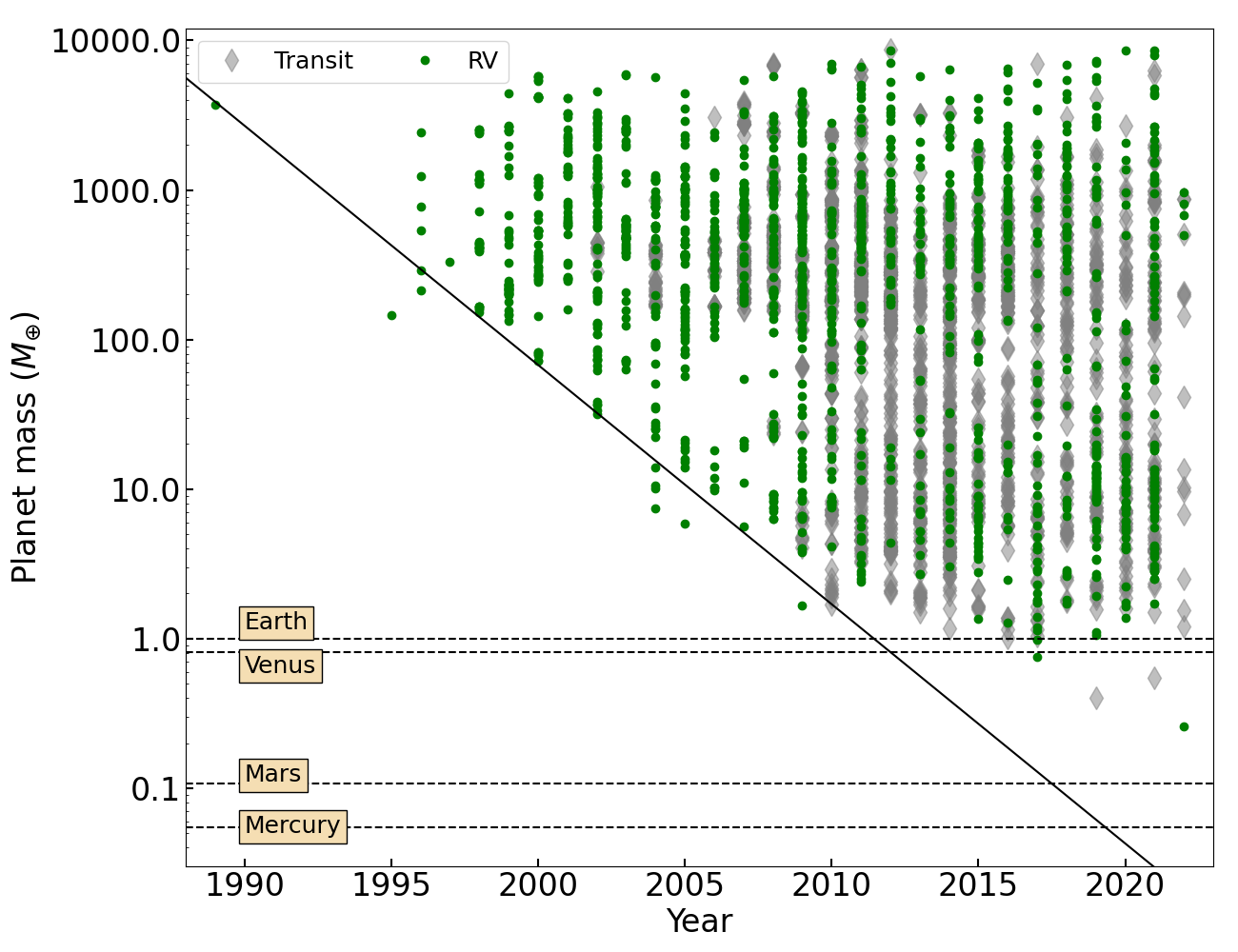}
\caption{Mass of known radial velocity planets (green filled circles) and transiting planets (grey filled diamonds) as a function of the discovery year. The horizontal dashed lines represent the mass of Earth, Venus, Mars and Mercury (from top to bottom). The black sloped line represents the plateaued sensitivity of exoplanet detection. }
\label{fig:massyear}
\end{figure}

Since the late 1980s \citep{Latham1989} there has been a logarithmic improvement in the smallest exoplanet mass detected, linearly with time; until about a decade ago 
(see Figure~\ref{fig:massyear}). 
Since sensitivity has plateaued: Corrections in stellar activity have struggled to improve exoplanet detection. With a few exceptions, for example,  \cite{faria_2022} announced the discovery of 0.26$\pm0.05$M$_{\oplus}$ planet orbiting around our nearest neighbour Proxima Centauri.

\cite{Langellier_2021} argued that a successful discovery of an exo-Earth requires a more sophisticated stellar variability model, longer-term instrumental stability and dense sampling of observations. This spawns a requirement of high-resolution spectra taken at a high signal-to-noise ratio with high cadence observations. Most spectrographs are already oversubscribed. Obtaining high signal-to-noise ratio a high cadence, high-resolution spectra is not sustainable.

In this paper, we argue that there is another way to help solve the stellar activity problem: to forecast when stellar minima happen. This will optimise the observing strategy of radial-velocity campaigns by focusing efforts on epochs when stars are quiet and activity can be more accurately modelled. In parallel, both {\it JWST} and {\it Ariel} are spacecraft with a limited lifespan. They require the most optimal and scientifically interesting targets. We argue that this philosophy should also apply to observing strategy: {\it JWST} and {\it Ariel} should observe at the most optimal and scientifically interesting times, hence the need to target low activity period, and thus the need for a stellar activity forecast shared amongst astronomers.

In \S 2 we demonstrate how different the sensitivity to small planets is affected by stellar activity by using publicly available solar radial-velocities obtained by HARPS-North at the different moments over a solar cycle. In \S 3, we present a simplified forecast and show how it can accurately predict stellar minima several years ahead. Then, we use publicly available data to forecast the stellar activity for {\it JWST}'s Early Release Science (ERS) and guaranteed time objects (GTO) in \S 4. Finally, \S 5 discusses the limitation of our current model, and the next steps we expect to take to improve our stellar activity forecast, including plans to make this model and its results publicly reachable.

\begin{figure*}
\includegraphics[width=0.495\textwidth]{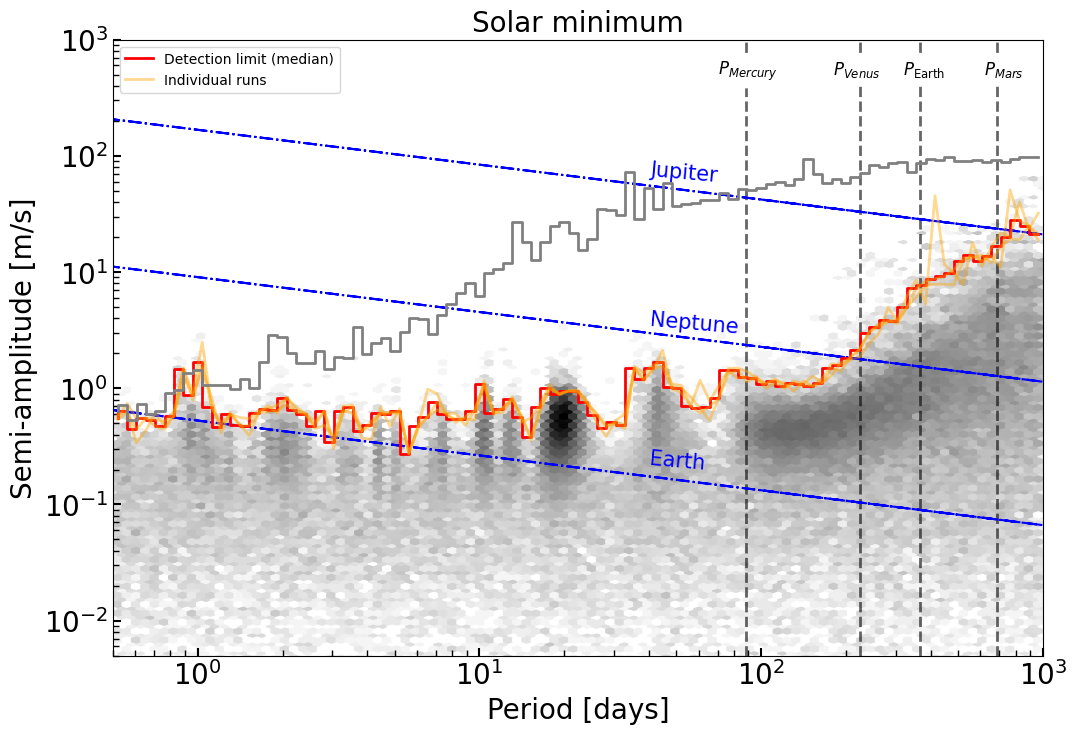}
\includegraphics[width=0.495\textwidth]{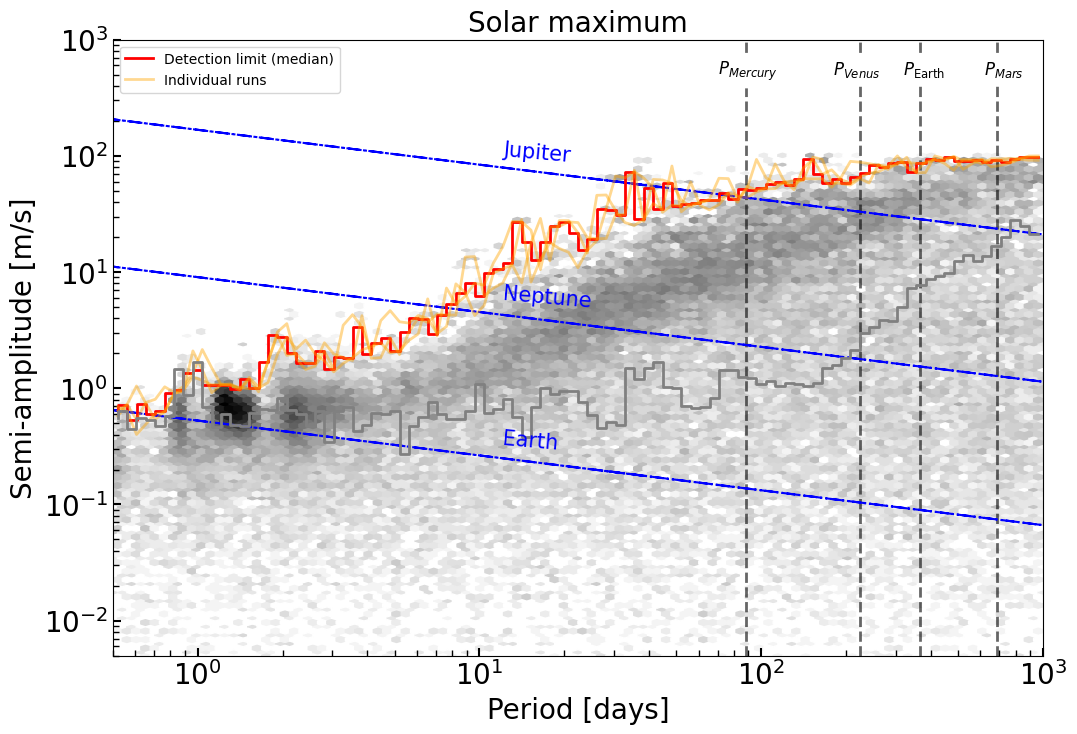}
\caption{The posterior distribution of radial velocity semi-amplitude plotted as a function of the orbital period to evaluate the detection limit for an exoplanet signal during solar minimum (left panel) and maximum (right panel). The calculated detection limit for an individual run is shown as orange lines. The solid red line shows the detection limit calculated from combined posterior samples combined. The detection limit improves by an order of magnitude during activity minima. For comparison, we overlay the detection limit during solar maximum as grey curve (left panel) and solar minimum in the right panel. Diagonal blue lines are anticipated signals of an Earth, Neptune and Jupiter mass planet.}
\label{fig:kimasim}
\end{figure*}

\section{Effect of the Sun's activity cycle on the detectability of planets}
A dominant characteristic of the Sun is the 11-year activity cycle as a result of  evolving magnetic field (\citealt{Balogh_2014}). 
During the activity cycle, the Sun undergoes a phase of weak and strong activity. During activity minima, the evolution of the dark and bright features on the solar surface is weak and vice versa during activity maxima (\citealt{Chatzistergos_2020, nandy_2021} references therein). 

Our goal in this section is to compare how activity minima and maxima of the Sun affect the detection of planets. We use the disc-integrated solar spectra provided by the solar telescope at the High Accuracy Radial-velocity Planet Searcher built for Northern hemisphere (HARPS-N, \citealt{phillips_2016, collier_2019})
We access the  solar radial velocities obtained from the HARPS-N spectrograph using the Data Analysis Center for Exoplanet (DACE) \footnote{\url{https://dace.unige.ch}}. We divide the data into solar maxima ($\rm JD \sim 2\,457\,200 - 2\,457\,400$) and minima ($\rm JD \sim 2\,458\,000 - 2\,458\,315$) based on a prediction by \cite{bhowmik_2018}. Each season contains 482 measurements, with a individual average precision of $1.07~\rm m\,s^{-1}$ \citep{dumusque_2021}. The HARPS-N solar data during activity minima ($\rm JD \sim 2\,458\,000 - 2\,458\,315$) contains 482 measurement. On the contrary during activity maxima between $\rm JD \sim 2\,457\,200 - 2\,457\,400$, nearly 1700 measurements were available. However, for simplicity 
we randomly chose 482 measurements during this period.

To compute detection limits, we use {\sc Kima} to fit Keplerian functions to the observed radial velocity data \citep{Faria_2018, standing_2021}. Based on a diffusive nested sampling algorithm, {\sc Kima} samples the posterior distribution of Keplerian parameters, including the number of planets, $N_{\mathrm{p}}$, within a system. $N_{\mathrm{p}}$ is sampled like any other free parameters \citep{brewer_2015, Faria_2018}. When there are no planets within a particular dataset, and if {\sc Kima} is forced to find some, it then returns a posterior that includes all solutions that are compatible with the data but remain formally undetected. This posterior can be used to compute a 99\% detection limit. We generally follow the procedure detailed in \citet{standing_2021} to analyse HARPS-N's solar radial velocities and produce detection limits.

Compared to \citet{standing_2021} who forced $N_{\mathrm{p}}=1$ to create detection limits, instead, we let $N_{\mathrm{p}}$ free with an upper limit of $N_{\mathrm{p}}=3$, which allows {\sc Kima} to explore the whole parameter space instead of converging on solar signals such as rotation. Like in \citet{standing_2021}, we set {\sc Kima}'s priors as log-uniform for the radial-velocity semi-amplitude ($K$), and the orbital period ($P$). We use a Kumaraswamy distribution  for eccentricity \citep{kumaraswamy_1980}. Our prior on eccentricity is $\alpha$=0.867 and $\beta$=3.00 \citep{kipping_2013}, while priors on the systemic velocity ($\gamma$) and the argument of periastron ($\omega$) are uniform. We perform three independent runs with {\sc Kima}, each resulting in $>20\,000$ effective posterior samples for data at solar minimum and repeat the procedure for solar maximum.
Since our fit finds $N_{\rm p} = 0$, all signals (planetary or stellar) remain statistically undetected and we can use the posterior distribution to compute a detection threshold.

We plot the combined posteriors of all three runs in Figure \ref{fig:kimasim}, shown as a log-log distribution of semi-amplitudes and the orbital periods during solar minimum (left panel) and maximum (right panel). We extract the detection limit for each of the three individual runs and plot them with yellow lines. The median of individual runs are depicted as orange lines. Doing so provides information about the uncertainty on the location of the detection limit. The orbital periods of terrestrial solar system planets are depicted as grey vertical lines. The green dotted-dash lines represent the expected semi-amplitudes for Jupiter, Neptune and Earth-mass planets as a function of orbital periods.

From Fig.~\ref{fig:kimasim}, we take the following lessons: the likelihood of detection of planets with  orbits $<1\rm ~day$ during activity minimum and maximum appears to be similar. However, for orbital periods $> 10~\rm days$ the detection limit is an order of magnitude greater at solar activity maximum compared to solar activity minimum. For instance, at P$_{\mathrm{orb}}\sim$100 days, the detection remains at Jupiter mass during activity maximum.

For solar minimum, the {\sc Kima} posterior shows an approximately flat detection limit until $P=180~\rm days$, which corresponds to the time span of the data we analyse. The posterior is denser around $P=22\rm~ days$, which likely corresponds to the solar rotation. Beyond $P=180~\rm days$, sensitivity decreases rapidly with orbital period, which is expected and also seen by surveys computing detection limits at orbital periods exceeding the survey's time span \citep[e.g.][]{howard_2010,mayor_2011,bonfils_2013,standing_2021}.
The main point to take from this analysis is that during solar maximum, at $P=180~\rm days$ the sensitivity to planets is $\sim 100 \times$ worse than during minimum. While some of the activity signals can be modelled, not having to correct means that modelling efforts can be focused on reducing stellar noise still present during minima instead.

\section{Construction of a simple stellar activity cycle forecast}

 We now detail two toy models forecast the stellar activity cycle, the first using spectroscopic data, and the second using photometry.
In principle, a forecast model should use all stellar activity indicators at once but for demonstration purposes we preferred isolating those. In  future publications, we plan to improve our forecast model and include other spectroscopic indices (such as $\rm H_\alpha$), and combine them with photometric time series, and UV/X-ray observations.

\subsection{Chromospheric activity cycle forecast}

To produce a first forecast we here focus only on the widely used Ca II H \& K lines. 
We follow the Mount Wilson Observatory procedure to investigate the stellar activity. We define two triangular pass bands  with a full-width half maxima of 1.09 ~\rm $\AA$ for the Ca II H \& K lines centred at 3968.47 ~\rm $\AA$ and 3933.664 ~\rm $\AA$. Additionally, we define two continuum bands with 20 ~\rm $\AA$ at 3901.070 ~\rm $\AA$ and 4001.070 ~\rm $\AA$ (\citealt{Noyes_1984}). The S-index is defined as 
\begin{equation}\label{eq:1}
   \mathrm{
S = \alpha \frac{H+K}{R+V}  
}  
\end{equation}
where H and K represent the total flux at the line centres, R and V represent the total flux in the continuum band pass, and $\alpha$ is a calibration constant adjusted at 2.3 \citep{duncan_1991}. 

The calcium S-index is related to the activity scale R$_{\mathrm{HK}}$. We define 
R$_{\mathrm{HK}}$ as the total flux at the stellar surface in the narrow band H and K lines normalised by the bolometric flux of the star. We converted the estimated S index to R$_{\mathrm{HK}}$ using the prescription given by \citet{oliveira_2018} and references therein.

We then carry out an analysis of the activity time series using a generalised Lomb-Scargle periodogram \citep{zech_2009} and compute false alarm probabilities to determine the significance of the observed signal. The individual point on the periodogram with N degrees of freedom and a frequency $\omega$ is calculated as
\begin{equation}
    {\rm power}= z(\omega) = \frac{N-3}{2} \times p({\omega})
\end{equation}
with:
\begin{equation}
    p(\omega) = \frac{\chi_o^2 - \chi^2(\omega)}{\chi^2_o},
\end{equation}
and where $\chi^2$ is the squared difference between the observed activity indicator $y_i$  and the model $y(t_i)$ is calculated as 
\begin{equation}
    \chi^2 = \sum \frac{[y_i - y(t_i)]}{\sigma_i^2}.
\end{equation}

To estimate the false alarm probability (FAP) associated with every point on the periodogram  we use 
\begin{equation}
    \mathrm{FAP} = 1-\left[1-\left(1-\frac{2P}{N-1}\right)^{\frac{N-3}{2}}\right]^M
\end{equation}
where P is the power of the significant period detected, N is the size of the dataset and M is the number of independent frequencies.

Using the rotation - activity cycle period relationship by \citealt{suarez_2016} (henceforth SM2016), we obtain a prior on the expected activity cycle period ($P_{\mathrm{cyc,prior}}$). The minimum frequencies for the periodogram are chosen to be $f_{\rm min} \sim \frac{1}{\alpha~ P_{\mathrm{cyc, prior}}}$. The factor $\alpha=1.5$ chosen arbitrarily.

We fit a simple sinusoidal model to the data with Markov Chain Monte Carlo using the \emph{emcee} package \citep{dan_foreman_mackey_2013}. We fit four parameters: the amplitude of the activity cycle (A$_{\mathrm{cyc}}$), period of the cycle (P$_{\mathrm{cyc}}$), the mean level of activity indicator ($\mu_{\mathrm{cyc}}$) and the epoch of minimum activity level.

\begin{figure}
\includegraphics[width=0.5\textwidth]{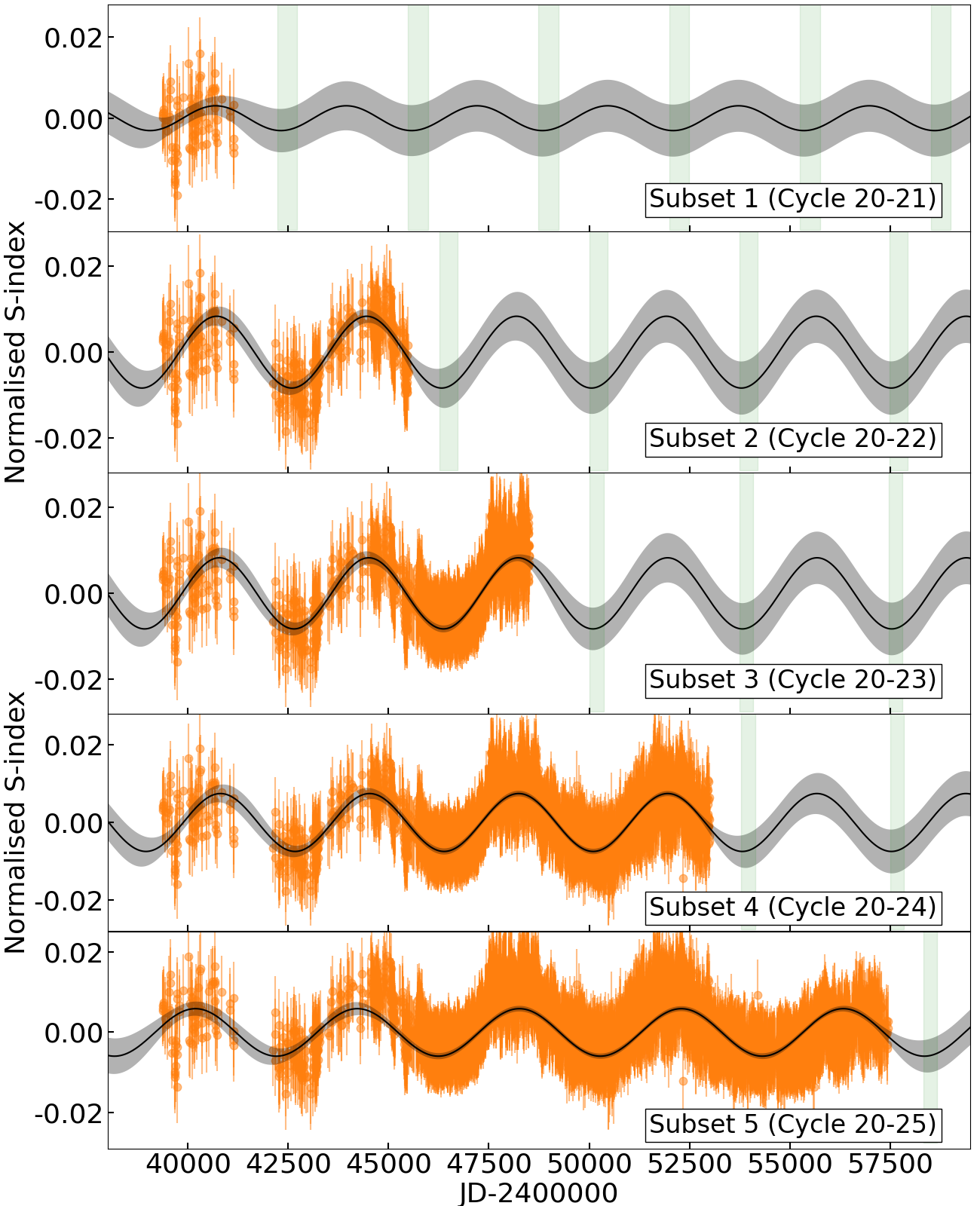}
\caption{The Mount Wilson solar Ca II H\&K index time series since 1966. Depicted in grey is the sinusoidal model predicting the solar activity cycle.
Each panel illustrates the improvement in the model with the inclusion of an additional subset of s-index data.}    
\label{fig:suncyc}
\end{figure}

\subsubsection{Case study 1 -  Solar activity cycle}

We use our Sun as a benchmark star to test our method of the forecasting activity cycle. We use the Mount Wilson solar s-index data \citet{egeland_2017} covering multiple activity cycles. 
Based on sunspot cycle numbers given in \citep{McIntosh_2020} and \citep{Hiremath_2022}, we divide the solar s-index data into five subsets to predict the cycle period and forecast the next activity minima. 
In Figure~\ref{fig:suncyc}, the s-index data of all the subsets are shown. We search for periodic signals in each of these subsets using a generalised Lomb Scargle periodogram and fit a sinusoidal model using MCMC. At each step we include the previous subset to improve the cycle prediction. 
Table~\ref{tab:sunres} provides the s-index cycle length and the forecasted next activity minima. The results indicate that the uncertainty on the forecasted minima epochs ranges between 0.5 and 0.8 years. However, in Appendix~\ref{apend_sind}, we treat each subset individually to obtain the cycle length and activity minima epochs. 

\begin{figure}
\includegraphics[width=0.5\textwidth]{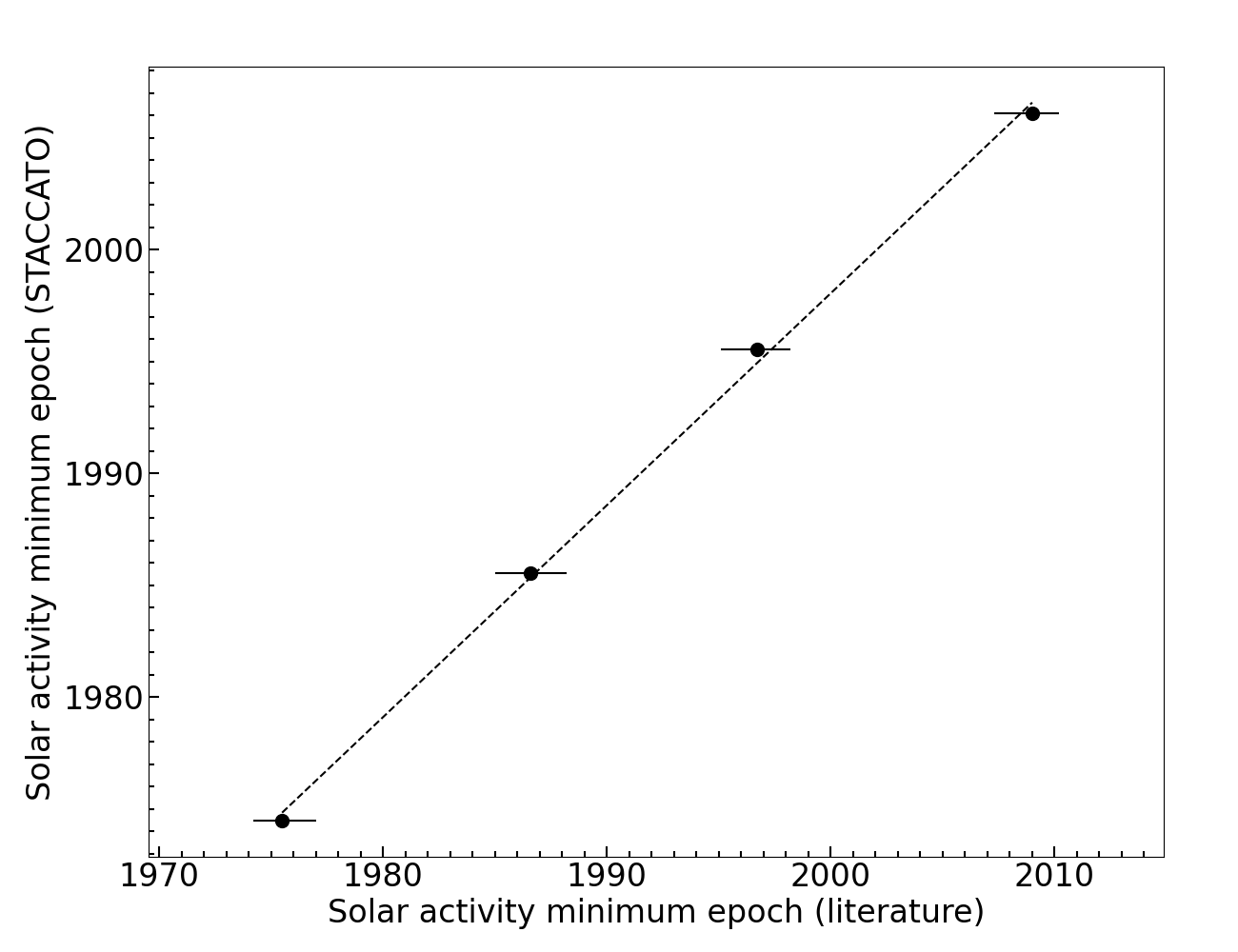}
\caption{ Regression plot between the STACCATO based solar activity minimum epoch versus the 
sunspot-less days reported by \citet{nandy_2011}. }
\label{fig:predvslit}
\end{figure}

In Figure~\ref{fig:predvslit}, we compare the STACATTO forecasted minima epochs with the epochs of sunspot-less days in the literature \citep{nandy_2011}. Our predictions are consistent with the literature value. 

\begin{table}
    \centering
    \begin{tabular}{ |c|c|c|c| } 
\hline
Subsets & Cycle length & Forecasted minima epoch \\
\hline
Subset 1 (Cycle 20-21) & 3259$\pm$151 & 2\,442\,121 $\pm$ 241d\\ 
Subset 2 (Cycle 21-22) & 3736$\pm$93 & 2\,446\,274 $\pm$ 227d \\ 
Subset 3 (Cycle 22-23) & 3725$\pm$25 & 2\,449\,917 $\pm$ 186d \\ 
Subset 4 (Cycle 23-24) & 3716$\pm$13 & 2\,453\,776 $\pm$ 174d \\ 
Subset 5 (Cycle 24-25) & 4038$\pm$14 & 2\,458\,204 $\pm$ 177d \\ 
\hline
\end{tabular}
    \caption{The length of activity cycle and the forecasted next activity minima epochs for individual subsets of solar s-index data.}
    \label{tab:sunres}
\end{table}

\subsubsection{Case study 2 - HD 51608, a star with an obvious activity cycle }

HD\,51608 is a $0.8\rm~M_{\odot}$ star known to host at least two Neptune-like planets with masses $12.7$ and $14.3~\rm M_{\oplus}$ orbiting every $14.1$ and $95.9~\rm days$, respectively \citep{udry_2017}. The host star is a moderately active star with $\log R'_{\mathrm{HK}} = -4.98\pm0.02$ and a stellar rotation period, $P_{\mathrm{rot}} = 40\pm4 ~\rm days$ \citep{udry_2017}. 
Using the SM2016 rotation-cycle period relationship, we predict the length of the magnetic activity cycle to be $\sim1556~\rm days$.

\begin{figure*}

\includegraphics[width=0.495\textwidth]{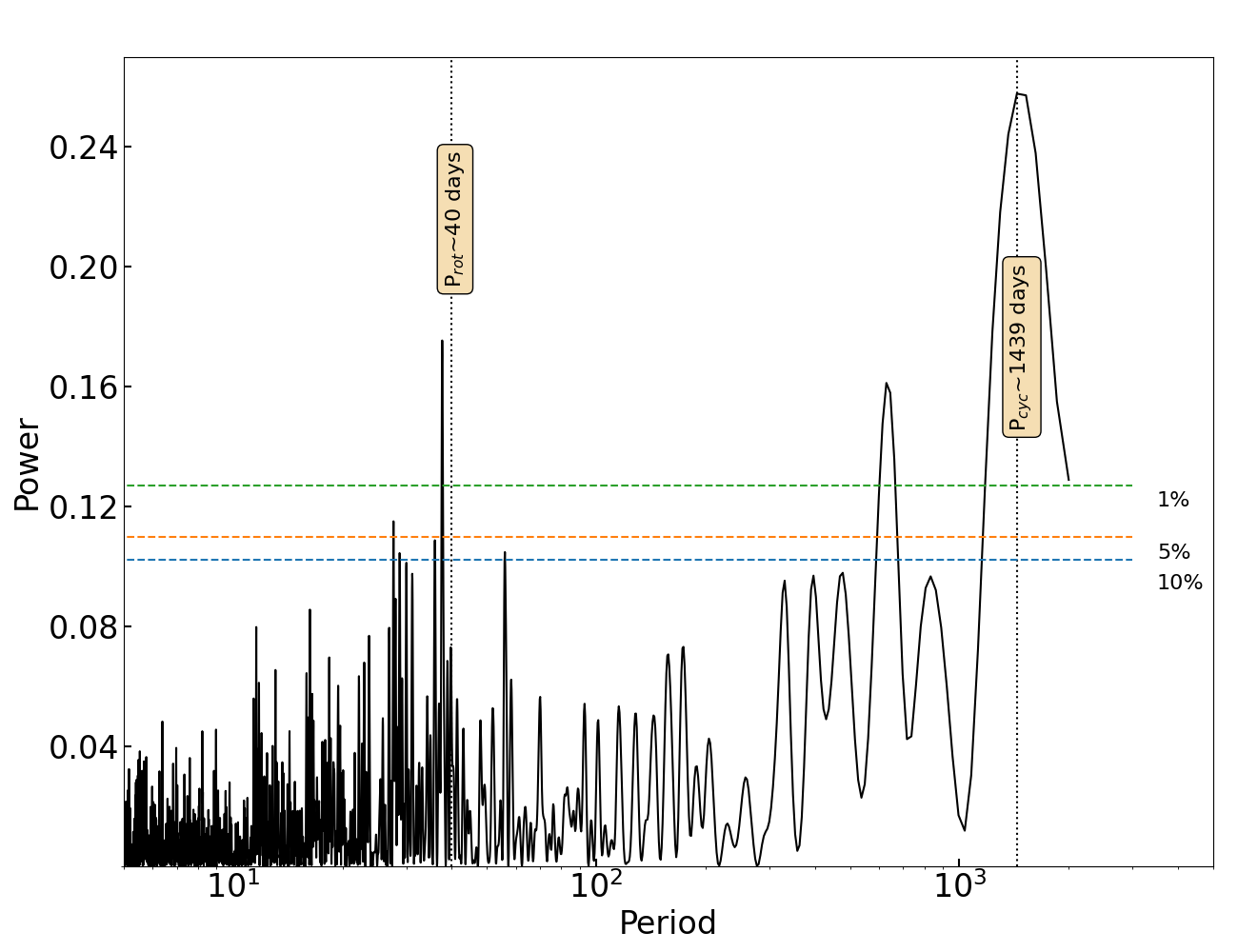}
\includegraphics[width=0.495\textwidth]{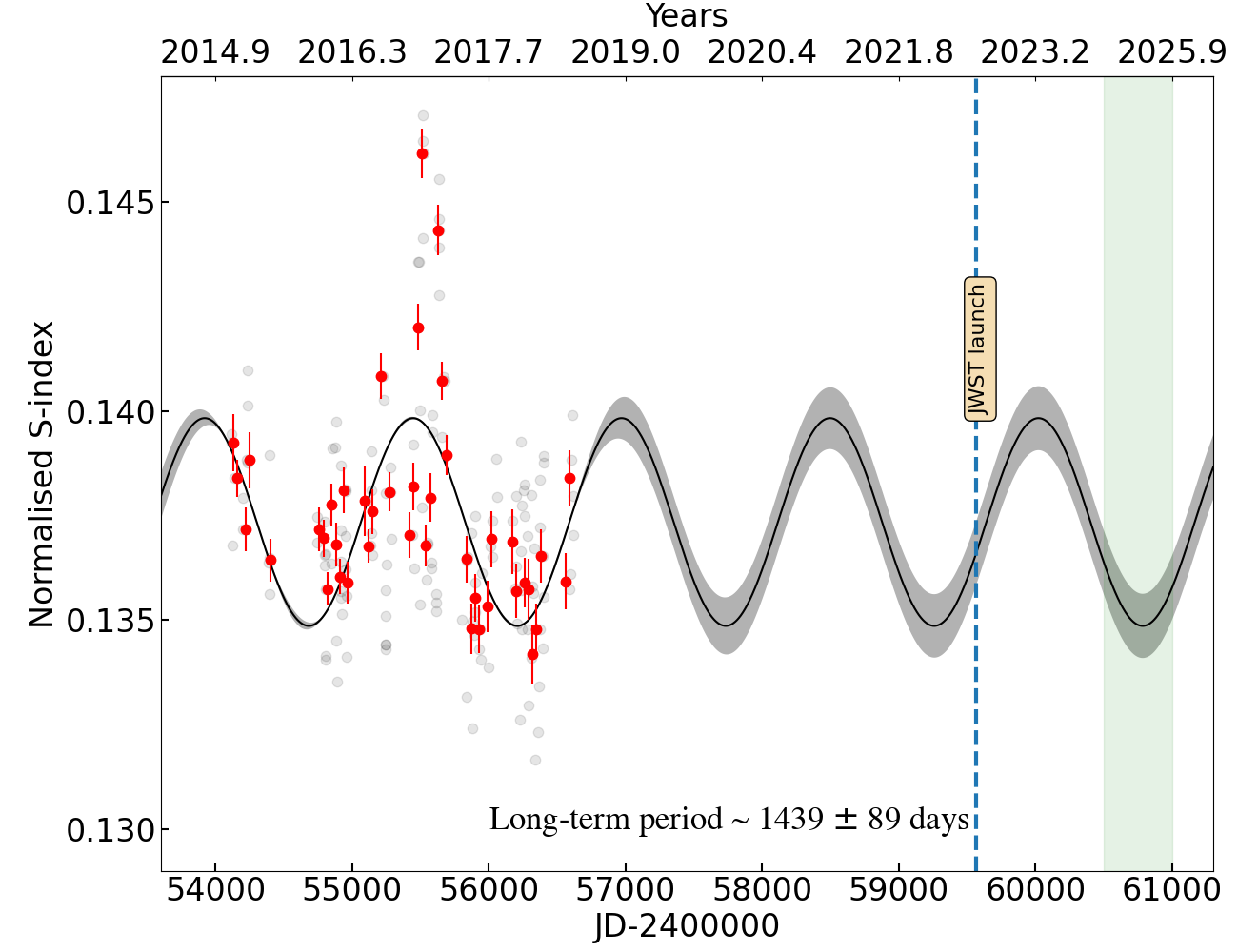}
\caption{Left panel: Periodogram of Ca II H\&K time series for HD 51608. The vertical dotted lines indicate the rotation period and the potential cycle period. While the horizontal dashed lines represent the 10, 5 and 1\% false alarm probability levels. Right panel: The phase fit for the long-term activity cycle with a period of 1439$\pm$89 days (3.9$\pm$0.2 years).Grey circles depict the HARPS s-index and red circles is monthly average HARPS s-index. The green block indicates the period of next activity minima. The blue dashed line indicates the phase of the activity cycle during the launch of JWST in December 2021. }
\label{fig:hd51608}
\end{figure*}

We collect public data from the HARPS ESO public archive. HD\,51608's HARPS data spans 2570 days. We used the already reduced spectra for HD\,51608 and performed our own measurement of the Ca II H\&K index using Equation \ref{eq:1}, producing a time series of activity indicators. In Figure~\ref{fig:hd51608} (left panel), we plot the periodogram with five significant peaks. The signal with period $P\sim38~\rm days$  corresponds to the rotation period of the stars ($P_{\mathrm{rot}} = 40\pm4 ~\rm days$). The longest period is likely induced by magnetic activity of the star.
The Ca II H\&K index data shows a very strong peak at $P\sim1400~\rm  days$. The modulation in Ca II H\&K shows a clear modulation with false alarm probabilities smaller than $0.1\%$.
On fitting the longest period signal to the activity index and subtracting it, except the $40~\rm day$ periods all the intermediate periods disappear. Furthermore, our MCMC run yields the cycle period of $1439\pm89 ~\rm days$ ($3.94\pm0.24~\rm years$) and forecasts the next epoch of activity minima $\rm JD \sim 2\,460\,485 - 2\,461\,011 $ (2024-06-23 to 2025-12-01) in Figure~\ref{fig:hd51608} (right panel). We interpret that previous observations cover at least two activity minima phases.

\subsubsection{Case study 3 - stellar activity cycle of LHS 1140}
LHS~1140 is an M4.5 star hosting at least two transiting planets with $6.48~\rm M_{\oplus}$ and $1.77~\rm M_{\oplus}$ orbiting every 24.7 and 3.77 days \citep{Dittmann_2017, Ment_2019}, respectively. LHS~1140 has been intensely monitored by instruments such as HARPS and ESPRESSO. The ESO archive provided in total 159 spectra spanning over 1368 days (116 HARPS spectra spanning over 671 days and 43 ESPRESSO spectra spanning over 303 days). The host star LHD~1140 is an inactive slow rotating star. Using HARPS and ESPRESSO radial velocity datasets \citet{lillo_2020} found the rotation period of LHS 1140 to be $\sim131~\rm days$. Using the SM2016 rotation-cycle period relationship, we predict the length of the magnetic activity cycle to be $\sim1850~\rm days$.

\begin{figure}
\includegraphics[width=0.5\textwidth]{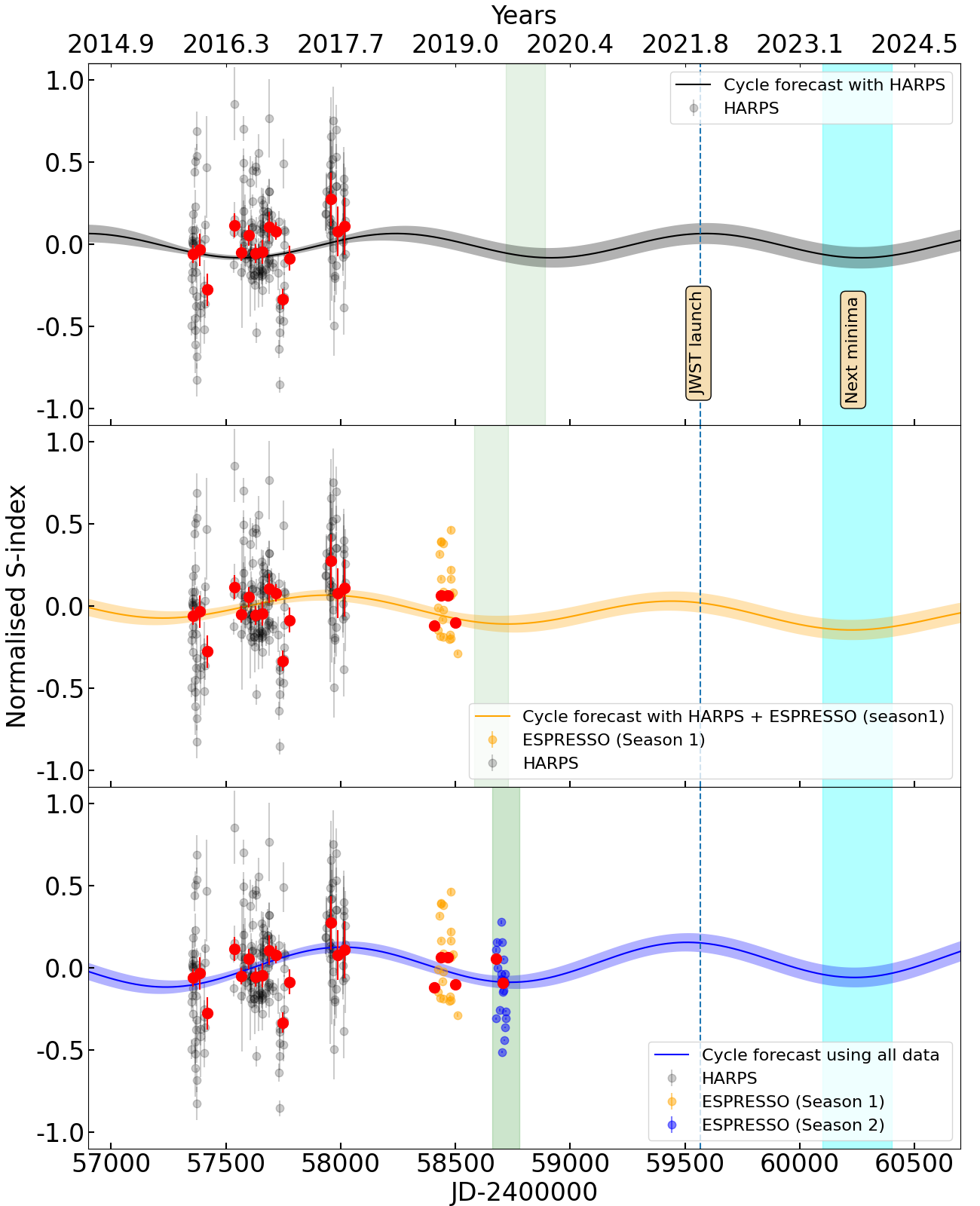}
\caption{The Ca II H\&K index time series for LHS 1140. Top panel: The activity indicator observed by HARPS along with the sinusoidal model depicting the long-term trend in the dataset. 
Middle and bottom panel: The HARPS and ESPRESSO data season-wise along with their sinusoidal models are depicted in grey, orange and blue, respectively. The red circles depict the monthly average  s-indices. The green region indicates the activity minima forecast from each observing run. The blue dashed line indicates the launch date of JWST in December 2021.  Depicted in cyan is the next forecasted activity minima for LHS 1140.  }
\label{fig:lhs1140}
\end{figure}

We begin by analysing the HARPS dataset obtained only between $\rm JD \sim 2\,457\,349 - 2\,458\,020$ (2015 November 22 - 2017 September 23), a  span of nearly two years. The Ca II S-indices for HARPS observations are depicted in Figure \ref{fig:lhs1140} top panel. Our best fit MCMC model yields a period of $\sim3.74 ~\rm\pm 0.48$ years with the next minimum activity epoch at $\rm JD = 2\,458\,805\pm 175$ which is represented with a grey sinusoidal curve in Fig. \ref{fig:lhs1140} (top panel). Although a large period trend is present in the data,  the nature of the signal still remains uncertain and might originate from an incompletely sampled activity cycle. 

We then add the first season of ESPRESSO observations (22 spectra from $\rm JD \sim 2\,458\,416 - 2\,458\,510$) and recalculate a forecast model to verify our earlier forecast of the next activity minima. We now find  $P = 4.13 \pm 0.36 $ years, and the next minimum at $\rm JD =2\,458\,712 \pm 132$, consistent with the first forecast. Finally, we add a second season of ESPRESSO data ($\rm JD = 2\,458\,674$ to  $2\,458\,719$) and this time, we obtain a cycle with $P = 4.04 \pm 0.28~\rm years$ with a predicted minimum at $\rm JD = 2\,458\,734 \pm 104$, consistent with the two previous attempts demonstrating that our toy model is predictive.

\subsection{Photometric activity cycle forecast}\label{subsec:phot}

For stars with sparse chromospheric/spectral observations, there exists publicly available photometric observations such as the All-Sky Automated Survey\footnote{\url{http://www.astrouw.edu.pl/asas/}} \citep[ASAS;][]{pojmanski_2002} that can be employed for forecasting stellar activity cycle.

\begin{figure}
\includegraphics[width=0.5\textwidth]{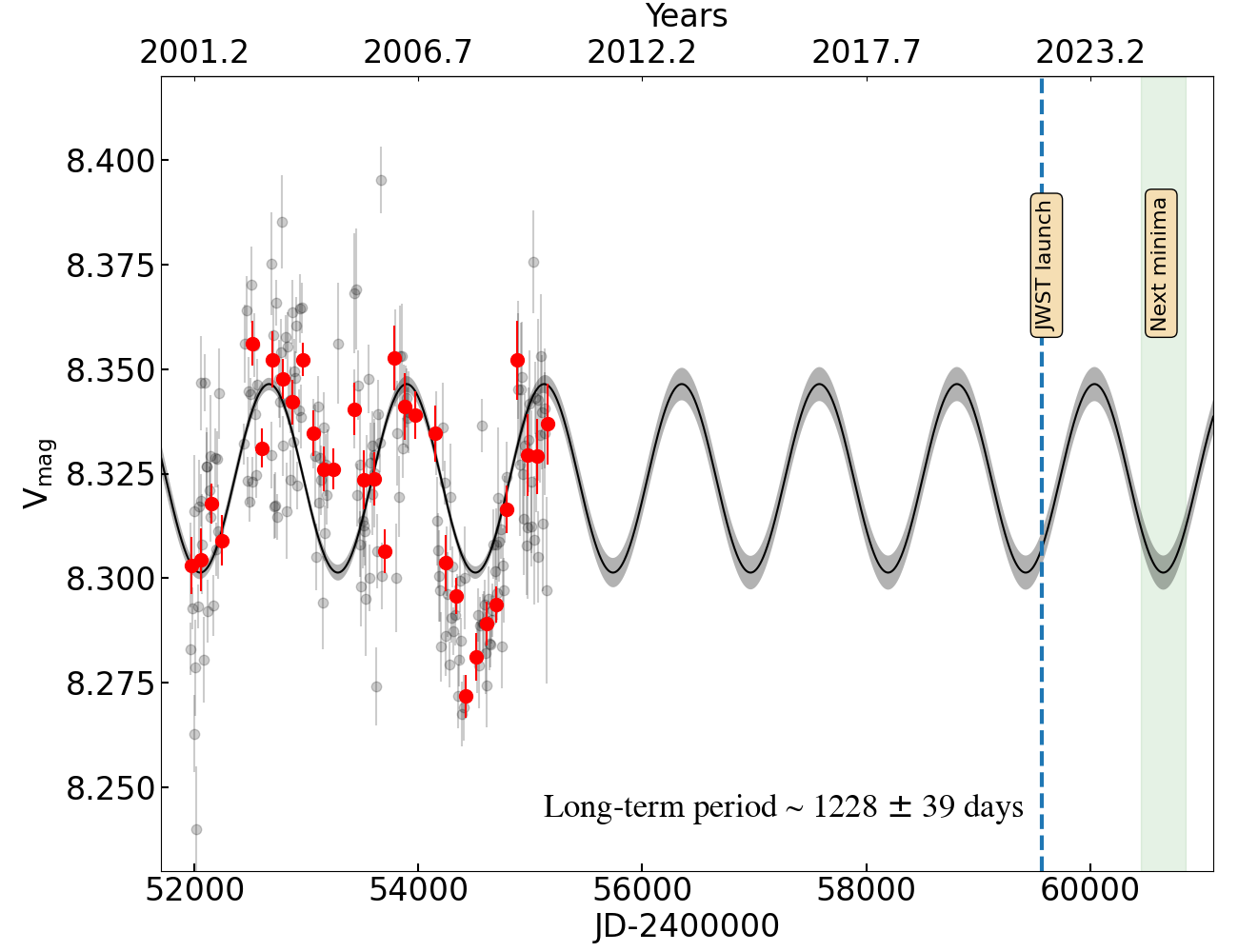}
\caption{ The long-term photometric cycle of HD 174429 with a period of 1228$\pm$39 days ($\sim$3.4 years). The black filled circles show the measure V$_{\mathrm{mag}}$ given by the publicly available ASAS database. The red circles are the quarterly binned V$_{\mathrm{mag}}$. The sinusoidal activity cycle model is depicted as a black curve. The blue dashed line indicates the launch of JWST. The green block depicts the next forecasted activity minima. }
\label{fig:hd174429}
\end{figure}

In our case study, we demonstrate such a forecast for a young active star HD 174429. This star is also called as PZ Telescopii (PZ Tel) a well known BY Draconis variable with a projected rotational velocity of 80 km s$^{-1}$ and rotation period of $\sim$ 0.943 $\pm$ 0.002 day \citep{maire_2016}. The ASAS data covers nearly 3000 days of observations. 
We search for periodic signals in the photometric data  by computing the power spectrum using a generalised Lomb Scargle periodogram. 
We are able to detect a long-term period of $1228\pm39~\rm days$. Figure~\ref{fig:hd174429} shows the phase fit for the bona fide long term signal. We fit this detected signal with a sinusoidal model using MCMC.  Depicted as a grey line is the predicted activity cycle with a length of $3.4\pm0.1~\rm years$.
Based on our analysis we think HD 174429 is currently transitioning from activity minima towards maxima. However, we forecast the next activity minima between $\rm JD \sim 2\,460\,436 - 2\,460\,832$ (2024-05-05 to 2025-06-05) and is expected to produce lower RMS in radial velocities and photometry, convenient for an eventual exoplanet detection survey focused on active stars.

\begin{figure}
\includegraphics[width=0.5\textwidth]{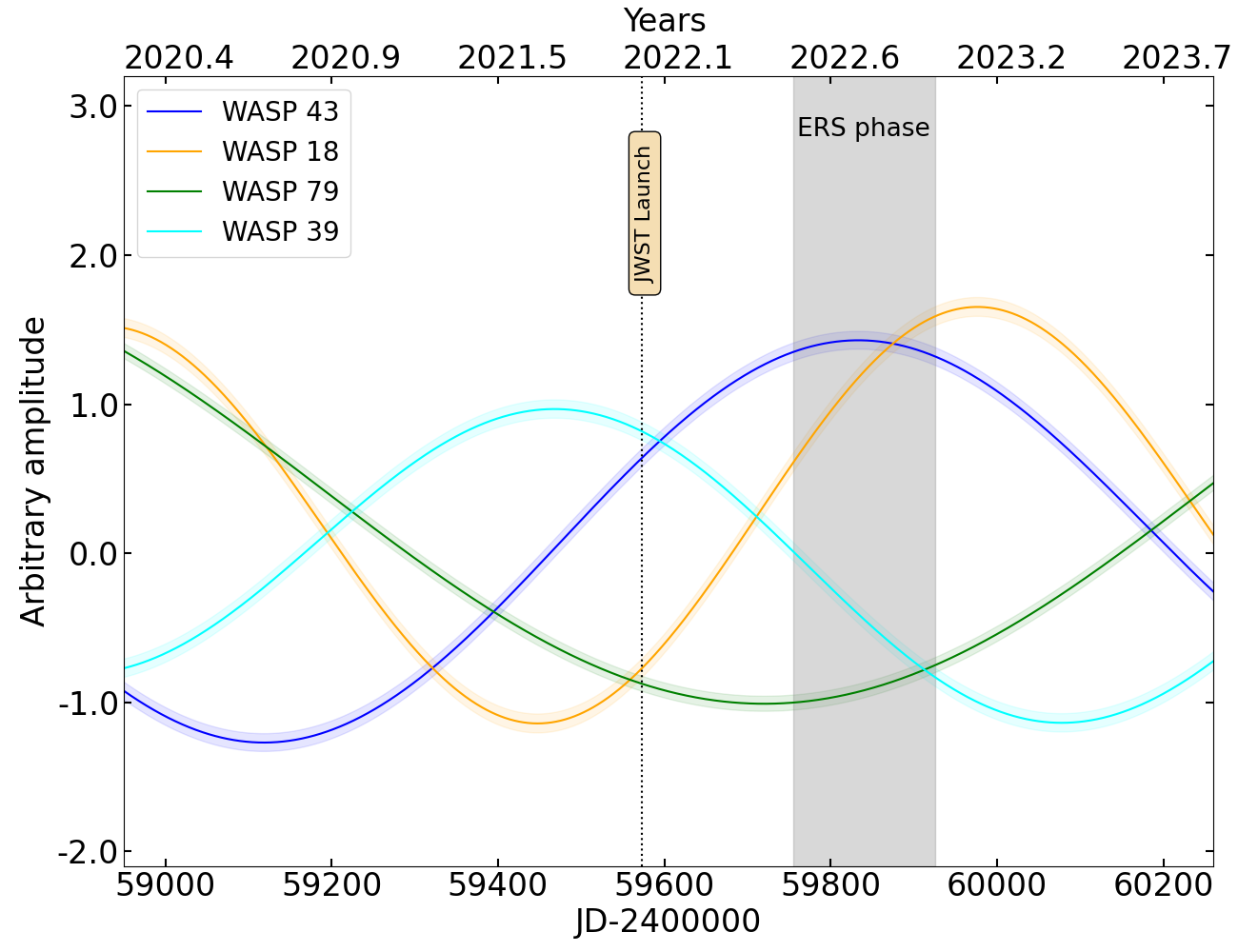}
\caption{Stellar activity forecast for WASP 43 (purple), WASP 18 (orange), WASP 79 (green) and WASP 39 (cyan). The vertical dotted line represents the launch of JWST and  the grey shaded region represents the JWST's ERS observation phase.  }
\label{fig:jwsters}
\end{figure}

\begin{figure*}
\includegraphics[width=1.05\textwidth]{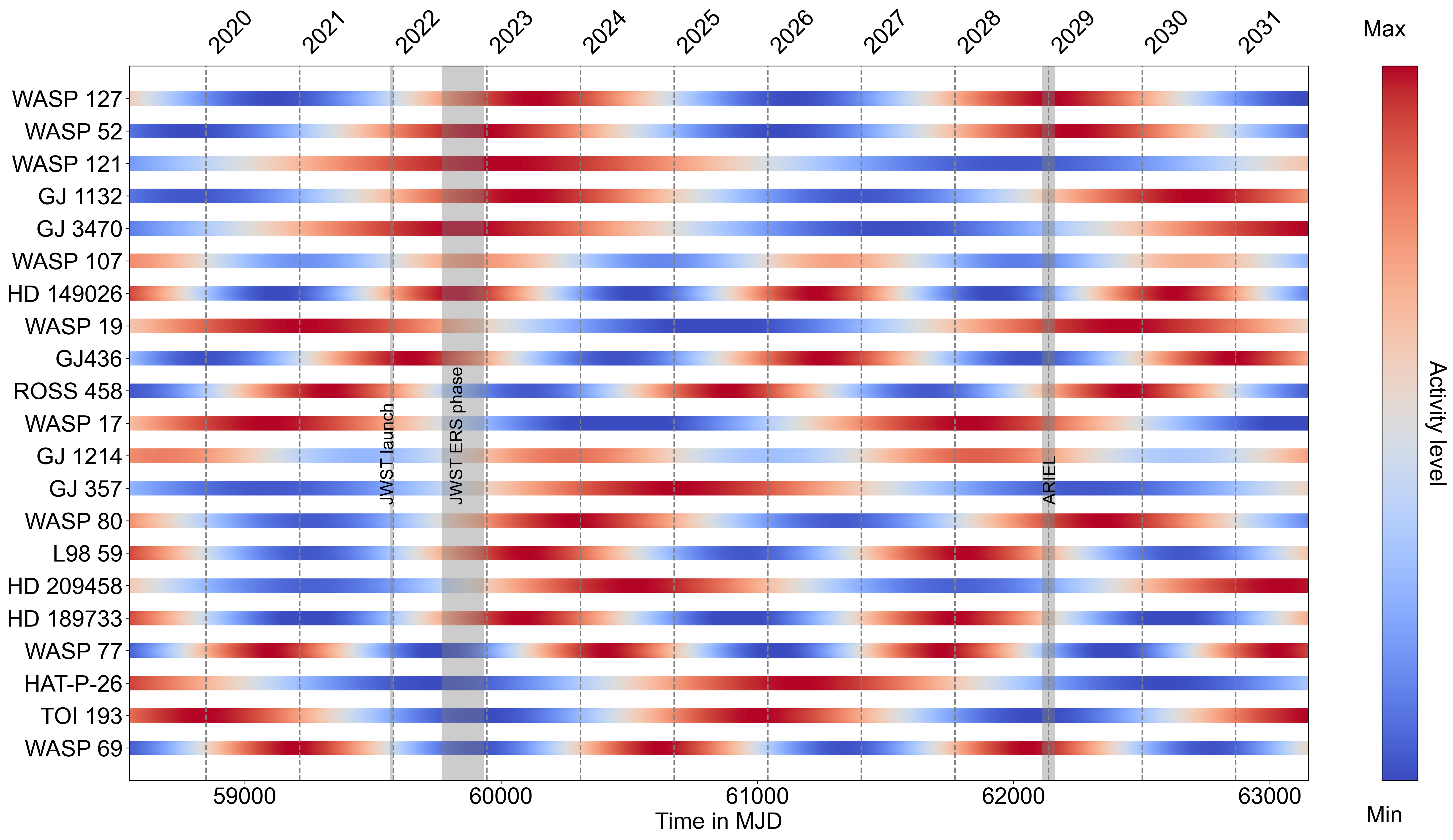}
\caption{Forecast for optimal observations of JWST's guaranteed time objects for the next ten years. The red and blue regions indicate the duration of activity maxima and minima, respectively. Depicted in grey from left to right are the launch of JWST, ERS phase and planned launch of ARIEL, respectively.}
\label{jwstplot}
\end{figure*}

\section{Application of a stellar activity forecast to planned JWST transit observations}

Recently {\it James Webb} launched. The spacecraft is due to begin science operation around summer 2022. There are two sets of exoplanetary systems we already know are planned to be observed, those selected by Guaranteed Time Observations (GTO) and those from the Early Release Science programme (ERS).

During the first months of {\it JWST} science operation, nearly 500 hours of director's time have been allocated for ERS of which 135 hours are dedicated for exoplanetary science \citep[80.2 hours are allocated for Transiting Exoplanet community and 54.8 hours for High contrast imaging;][]{Bean2018}\footnote{\url{https://www.stsci.edu/jwst/science-execution/approved-ers-programs}}. The objective of the ERS program will be a) to provide quick first-look data for wide community b) performance verification of instruments c) prepare the community for open time proposals.

The transiting exoplanet ERS community will exercise the time series modes of all four instruments on board {\it JWST}. Three targets are to be observed as part of {\it JWST}'s ERS program are  WASP-43, WASP-18 and WASP-79. 

WASP-43 is the most active star in the {\it JWST} ERS sample with a 
 measured $\log R'_{\mathrm{HK}} = -4.17\pm0.10$ \citep{staab_2017} and rotation period of $15.6~\rm days$ \citep{hellier_2011}.The $\log R'_{\mathrm{HK}}$ of WASP-18 is -5.43 \citep{knutson_2010} with rotation period of $5.6~\rm days$ \citep{Hellier_2009}. The rotation period of WASP-79 is $4.0\pm0.8~\rm day$ \citep{smalley_2012}. Our analysis of HARPS archival data for WASP-79 yields $\log R'_{\mathrm{HK}}= -4.68\pm0.10$ (see Appendix \ref{apend_rhk} for details).

Using the SM2016 relationship we estimate the activity cycle lengths to be $\sim1390~\rm days$ for WASP-43, $\sim 1117~\rm days$ for WASP-18 and  $\sim 1870 ~\rm days$ for WASP-79. Initially, WASP-39 was planned to be observed however, a delayed launch for {\it JWST} forced a switch to WASP-79. Using these estimated cycle lengths and combining them to archival photometric data to find the phase of the activity cycle, using the method described in section~\ref{subsec:phot}. Our results are shown in Figure~\ref{fig:jwsters}. Based on the activity forecast, WASP-79 (depicted in green) is better suited to be observed between June and December 2022 due to its predicted activity minima phase which corresponds well to its planned ERS observations. However, for WASP-43b (purple) and WASP-18b (purple), the current period of observation will, unfortunately, coincide with the highest part of their host star's magnetic cycle. Any  exoplanetary atmospheric observations carried out during this phase will have to be interpreted cautiously due to likelier stellar activity contamination. This is particularly important since those ERS observations are meant to inform the community which {\it JWST} observation modes are best suited.

Furthermore, we identify 21 unique exoplanet hosts to be observed spectroscopically with {\it JWST} as part of its GTO, to characterise the atmosphere of these planets. Most of these targets have been monitored spectroscopically and photometrically with public data covering a large timescale. They also have a well-defined rotation period meaning we can once more use Suarez\,Mascare\~no relation. We forecast the evolution of the activity cycles over the next 10 years. In Figure~\ref{jwstplot}, we show the optimal observing time for {\it JWST}'s GTO targets over the next 10 years. The redder the region indicates that the star is in higher activity state (activity maxima) leading to a more likely contamination and less optimal for observation. The ideal observing times for each of the star is indicated in blue, during low activity state (activity minima). Currently, WASP-127, WASP-52, WASP-121, GJ\,1132, GJ\,3470, WASP-107 and HD\,149026 (from the top of Fig.~\ref{jwstplot}) are in  a high activity state (maxima). 
We forecast at least 6 more targets heading towards activity maximum in the next 6-8 months. The GTO will be scheduled amongst the first observations carried out by {\it JWST}. Indicating potential contamination of observations due to high levels of stellar activity. 

Within the GTO lists, the planet more ideally observable for transmission spectroscopy are at the bottom of Fig.~\ref{jwstplot} are and WASP-69, TOI-193, HAT-P-26, WASP-77, WASP-17 and ROSS 458. We note that these forecasts are indicative only at this stage since they are based on a limited number of repeat visits. With additional activity indices/observations, and a more dedicated monitoring, there is scope to improve the forecasting accuracy.

\section{Conclusions and future developments}

First, we show how activity cycle affects the detection efficiency of planets in the solar case. The detection of a low mass planet decreases by an order of magnitude during solar activity maximum. Hence the successful discovery of Earth analogues and their characterisation would be improved with a more optimal observing strategy based on a forecast of stellar activity. 

We then present a novel method to characterise the magnetic activity cycle of a few stars using archival data and use this to accurately forecast when the next stellar minimum will be, years ahead. This allows to a more optimal schedule of observations, be they for exoplanet detection or for atmospheric transmission spectroscopy. Even when the amount of available archival data is limited, we can use them in combination to the relationship between the rotation period and activity cycle periods from SM2016 to obtain the phase of the magnetic activity cycle for any star and thus produce a first forecast of the next minimum. Like for any forecast we expect that its accuracy would increase with increasing amount of data and proximity to the target date (here stellar minima). We note however that the solar maxima is only predicted to about $\pm 0.5-1~\rm year$ ($\sim 5-10\%$; \citealt{kitiashvili2016data,chowdhury2021prediction}), a precision that other stars are likely to match.

Our current method for forecasting stellar activity cycles has a number of simplifications. For instance it assumes the SM2016 rotation/magnetic period relation holds true for all stars, and relies on these stars having regular activity cycles. Despite its simplicity, our method nonetheless could produce activity cycle forecasts for stars from G to M types surprisingly well.

Following this proof of concept, we envisage several obvious subsequent steps to improve our forecasting model and broadcast its results, which we expect to detail and test in forthcoming papers.

\begin{enumerate}
    \item Whilst, Ca II H and K lines have been the most widely used chromospheric indicator, they are of decreasing use for low-mass K and M stars. that are redder and intrinsically fainter than Sun-like stars producing severely reduced signal-to-noise ratios.  Therefore, other activity sensitive lines such as H$\alpha$, the Na I doublets, and the He I lines are obvious elements of an improved and more widely applicable forecasting model. 
    \item So far we have only used photometry sparsely, however photometric measurement are usually more easily available (i.e. {\it TESS } and {\it PLATO}) and are a natural next step to improve the accuracy and reach for our model. 
    \item Our model only applies sine curves to the amplitude of the variability. An improved model should also extract information from the periodic variability of the scatter of spectroscopic and photometric measurements, since rotational variability is more pronounced during activity maxima.
    \item Our model uses literature values for the rotation period. An improved model should measure that rotation period directly and apply the Suarez\,Mascare\~no relation simultaneously.
    \item A useful forecast is an available forecast. We also plan to open a searchable web service providing forecast for all stars data can be gathered for. Forecasts will regularly be updated as new data become periodically available.
    \item For stars with multi-year activity time series applying deep-learning algorithms might helps forecast the amplitude of upcoming activity cycles. 
\end{enumerate}

While forecast can be effective in increasing the efficiency of detection of exoplanets as well as their chemical species on ongoing projects, the method has potential for application in other uncharted areas of exoplanetary science. Two possible prospective applications of such forecast are 
\begin{enumerate}
    \item A transition in radial velocity survey from inactive to active stars: Because of their age, young active stars are ideal candidates providing a novel environment within which the consequence of planet formation can be witnessed, enabling us to deepen our understanding of the evolution of planetary systems. Due to challenges posed by stellar activity, ongoing RV surveys do not include young active stars. Activity cycle-based forecast of the optimal observing time of young active stars will produce lower root mean square scatter in radial velocity. Hence, increasing the probability of detecting radial velocity planets around these stars.
    \item Systematic detection of auroras in exoplanets: Since auroras are generated by solar events like bright flare and coronal mass ejection, we see more auroral activity on Earth during solar maxima. Hence, forecasting the activity maxima for host stars could present a good prospect for auroral detection in exoplanets.
\end{enumerate}

Collecting all these stellar activity data is also an opportunity to open new lines of enquiry, extending the Mount Wilson survey to thousands of systems. With this information it might be possible to establish a correlation between the photometric cycle and chromospheric activity level. This relationship between the basal chromospheric activity level and the photometric activity cycle is crucial for stars with low cadence data and without any measured rotation period. 
Secondly, only about 30 solar cycles have been properly recorded in human history and the probability of events such as the Maunder minimum are unknown. Cycles collected for thousands of stars might reveal how regularly such patterns occur.

\section*{Data Availability}

The data we used is available for download at the ESO public archive\footnote{\url{http://archive.eso.org/eso/eso_archive_main.html}} and the Data \& Analysis Center for Exoplanets (DACE), which is a facility based at the University of Geneva \footnote{\url{https://dace.unige.ch/observationSearch/?observationType=[\%22solarSpectroscopy\%22]}}.

\section*{Acknowledgements}

We would like to thank our reviewer Alejandro Suarez-Mascareno for his comments which helped improve the manuscript. 
This research is based in part on data obtaine dwith ESO telescopes at the La Silla Paranal observatory under programme IDs 072.C-0488(E), 091.C-0936(A), 0100.C-0884(A), 0102.C-0294(A), 0103.C-0219(A), 0104.C-0316(A), 183.C-0972(A), 191.C-0873(A), 192.C-0852(A), and 198.C-0838(A). This paper includes data collected by Mount Wilson survey and All-sky Automated Survey. 
This work was supported by a Leverhulme Trust Research Project Grant (grant number RPG-2018-418).



\bibliographystyle{mnras}
\bibliography{paper} 




\appendix

\section{Method description for \texorpdfstring{$\log R'_{\mathrm{HK}}$ C}   calculation}\label{apend_rhk}

We used the method described in \citep{schroeder_2009} for removing the colour dependence and contribution of photospheric component in the activity indicator. The S index is transformed into $R_{\mathrm{HK}}$ as follows:

\begin{equation}
\mathrm{
    R_{HK} = 1.34 \times 10^{-4} ~C_{cf} ~S.
    }
\end{equation}

For main sequence stars with $0.3\leq\mathrm{B-V}\leq 1.6$, the conversion factor $\mathrm{C_{cf}}$ is  given by \citep{middelkoop_1982, rutten_1984}

\begin{equation}
\mathrm{
    log ~C_{cf} = 0.25(B-V)^3 ~-~ 1.33(B-V)^2 ~+~ 0.43(B-V)~+~0.24
}
\end{equation}

The correction for the photospheric contribution to the Ca II line core fluxes were applied using \citep{Noyes_1984} relationship

\begin{equation}
    \mathrm{
    log~R_{phot} = -4.898 + 1.918(B-V)^2 - 2.893(B-V)^3
    }
\end{equation}

The photospheric corrected  $R'_{\mathrm{HK}}$ is given by 
\begin{equation}
    \mathrm{
     R'_{HK} = R_{HK} - R_{phot} 
    }
\end{equation}

\section{Solar s-index cycle}\label{apend_sind}

\begin{figure}
\includegraphics[width=0.48\textwidth]{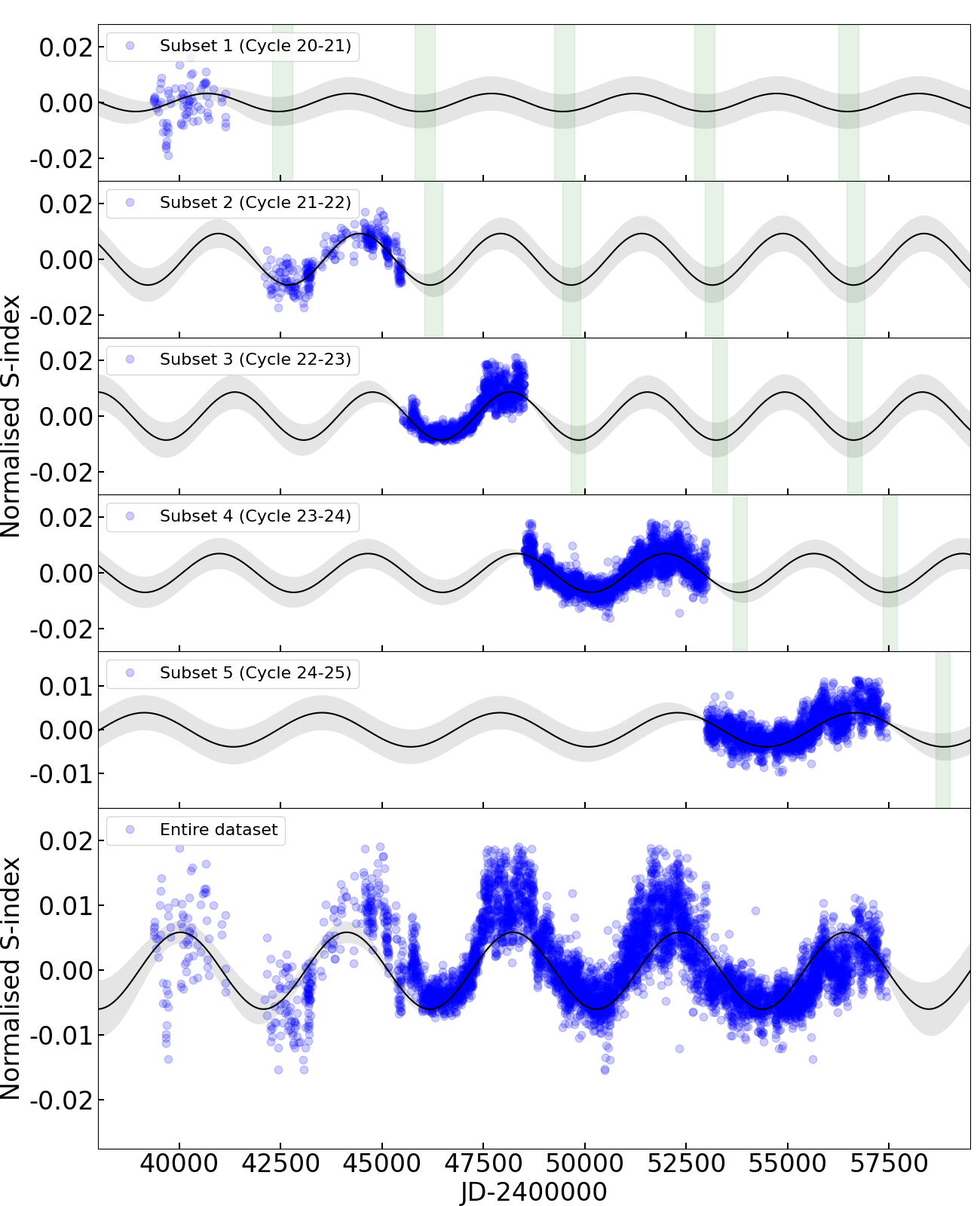}
\caption{The Mount Wilson solar Ca II H\&K index time series since 1966. Depicted in grey is the sinusoidal model predicting the solar activity cycle.
Each panel illustrates the model for an individual subset of s-index data. The combined subsets along with the improved forecasted model are shown in the bottom panel.}
\label{fig:sunappend}
\end{figure}

The solar s-index data \citep{egeland_2017} is divided into subsets and modelled individually to obtain the activity cycle lengths and activity minimum epochs (Figure~\ref{fig:sunappend}). The predicted minimum epochs and lengths are consistent within 0.5-1 year.

\end{document}